\documentclass[aip,rsi,reprint]{revtex4-1} 

\usepackage{amsmath}
\usepackage{graphicx}

\graphicspath{{Figs/}}

\draft 

\begin{document}


\title{A fast pneumatic sample-shuttle with attenuated shocks.} 



\author{Valerio Biancalana}
\affiliation{DIISM and CSC University of Siena; CNISM}

\author{Yordanka Dancheva}

\author{Leonardo Stiaccini}
\affiliation{DSFTA and CSC University of Siena}


\date{\today}

\begin{abstract}
\footnotesize
We describe a home-built pneumatic shuttle suitable for the fast displacement of samples in the vicinity of a highly sensitive atomic magnetometer. The samples are magnetized at 1~T using a Halbach assembly of magnets. The device enables the  remote detection of free induction decay in ultra-low-field and zero-field NMR experiments, in relaxometric measurements and in other applications involving the displacement of magnetized samples within time intervals as short as a few tens of milliseconds. Other possible applications of fast sample shuttling exist in radiological studies, where samples have to be irradiated and then analyzed in a {\it cold} environment. 
\end{abstract}



\maketitle 

\section{Introduction}
\label{sec:introduction}
Several kinds of measurements require the fast displacement and an accurate positioning of prepared samples. Samples need to be displaced from a {\it preparation} region to a {\it measurement} region, which coincides with the vicinity to a sensor head. This kind of remote detection technique has been used in relaxometry \cite{kimmich_pnmrs_04, swanson_jmr_93, wagner_jmr_99}, dynamical nuclear polarization experiments \cite{reese_amr_08, krahn_pccp_10}, and in low-field-,  ultra-low field-, and (near-)zero-field- NMR spectroscopy \cite{weitekamp_prl_83, bielecki_rsi_86, kerwood_jmr_87, bevilacqua_arnmrs_13}. Sample shuttles are also used in radiological analyses \cite{ismail_jammc_10}.
Liquid and gaseous samples can be efficiently displaced by means of pumps and pipelines \cite{bevilacqua_jmr_09}, while solid samples (as well as sealed liquid samples) can effectively be displaced by means of so-called {\it rabbit} systems.

In this note we describe a home-built pneumatic shuttle capable of moving plastic cartridges containing solid or liquid samples of a few ml in volume along a tube a couple of meters in length. The transit time is as short as a few tens of milliseconds. The setup is automatized to facilitate repeated measurements. A pneumatic method is applied  to decelerate the sample, which helps to reduce the shock intensity and to prevent bouncing: important issues for reproducible and precise sample positioning. Specifically, the setup is developed to be coupled with an atomic magnetometer \cite{bevilacqua_pra_12, belfi_josa_09}, which  measures the sample's magnetization and its time-evolution. Thus the whole setup enables remote detection of free-induction-decay or relaxometric signals \cite{bevilacqua_jmr_09, belfi_rsi_10}. 

An obvious trade-off exists, demanding  a compromise between short displacement time, long distance, and small acceleration and shocks. In addition, repeated measurements (such as tens, hundreds or even thousands of cycles) must be feasible, meaning that materials must be selected carefully in order to reduce surface abrasion and cartridge or tube deterioration.

\section {Shuttle}
The sample is enclosed in a plastic cartridge which moves through a transparent plastic tube. 
The tube is 2~m in length,  while the external and the internal diameters are $d_{\mbox{ext}}=$25~mm  and $d_{\mbox{int}}=$19~mm, respectively. The cartridges are lathed to fit $d_{\mbox{int}}$ with 0.15~mm tolerance.

\begin{figure} [htbp]\centering
 \includegraphics [width=8cm] {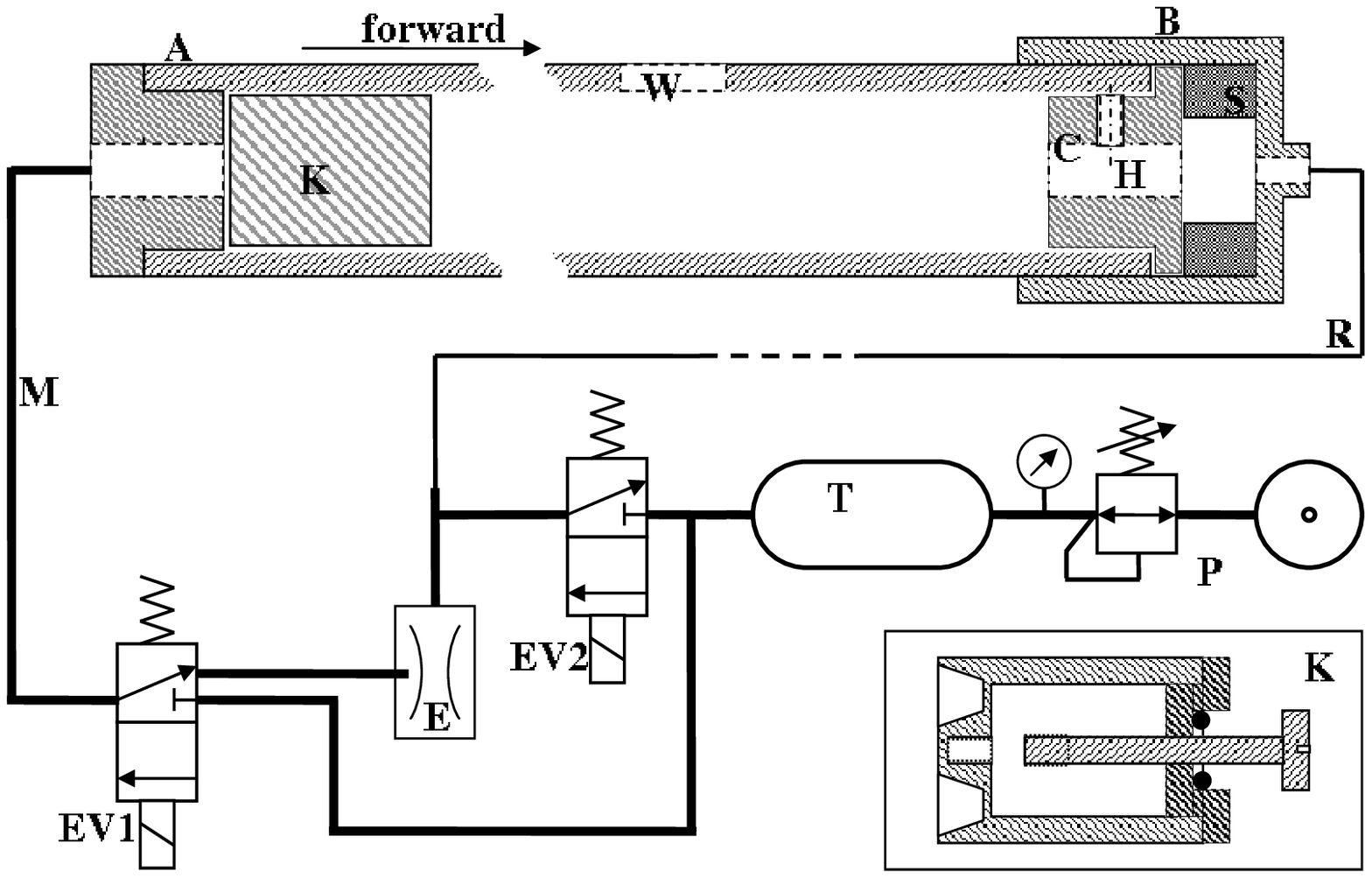}
 \caption{Schematic of the pneumatic shuttle setup and of its connections (not to scale). P=air compressor, pressure regulator, and manometer; T=reservoir; EV1 forward electro valve; EV2=backward electro valve; E=ejector; A=premagnetization end; B=measurement end; W=side air-escape window; C=shock absorbing cylinder; S=sorbothane ring; K=cartridge; H=calibrated air flux control; M, R = air connection pipelines. The inset represents the physical structure of the cartridge K, which is kept sealed by means of a plastic screw.
 \label{fig:setup}}%
\end{figure}

Fig.\ref{fig:setup} represents the tube and the schematics of the air connections.
Both  fast forward displacement and  (slower) backward displacement are actuated with an air connection in  end A of the tube, while a thin pipeline (R) feeds  end B with a small air flux, necessary to start the backward motion.

The forward motion of the cartridge (K) is activated with a 687~Nl/min electro valve (EV1), through a short connection (M: 50~cm in length, 8~mm in diameter) between  end A and a reservoir (T), whose 3~litre volume far exceeds the tube volume. A large side hole (W) drilled at about 10 centimeters from  end B, lets the air  escape during the fast forward displacement. After passing W, the cartridge compresses the small amount of air present in the  WB segment of the tube. During this compression, that small amount of air escapes toward the pipeline R, passing through an air-flux control (H), which is adjustable by means of a side-screw. The size of H is calibrated as so to minimize the energy of the final collision. In other terms, the cartridge is pneumatically decelerated during the WB displacement. This feature reduces the collision shock and greatly relaxes the requirements in  terms of the mechanical strength of the cartridge.

The cylinder (C) is slightly lighter than the cartridge and absorbs the small amount of residual energy (if any),  losing it on a Sorbothane ring (S). After some calibration work (selection of the size of H) the deceleration system described became reliable and reproducible, avoiding significant shocks due to hard K-C collisions and preventing the cartridge from bouncing at end B.

In cycles of measurements, after the detection, the cartridge (K) must be moved back to  end A. For this purpose a smaller (225~Nl/min) electro valve (EV2) is actuated, providing both a small air flux in R and a larger flux to feed the ejector (E). The air flux passing through R and C pushes K toward  W. Once  K has reached W,  the depression produced  by the ejector pulls the cartridge back to A, making the system ready for the next cycle. Typical shuttling times are 50~ms for the forward displacement and 1200~ms for the backward one, for a tube of 2~m in length. The ejector is a home-built, adjustable, aluminium device, the designs for which are available among the supplemental material.

The cycle timing is controlled and verified by means of a data acquisition card driven by a LabView code, which will be integrated into the program controlling the magnetometric setup. Two digital outputs are used to activate the electro valves EV1 and EV2 through  opto-coupled MOSFETs, while the analogue signal produced by two photodiodes, detecting the cartridge's passages in two adjustable positions, is analyzed to infer the speed and acceleration of the sample. 
 
The filling pressure of the tank (T) is regulated and stabilized to 0.6~MPa, which is a safe value for the tube used, according to the empirical expression for maximum pressure provided by the supplier as:
$ p_{\mbox{max}} \cong 2.5 \times (d_{\mbox{ext}}/{d_{\mbox{int}}-1})$~MPa~$\cong 0.8$~MPa.

\begin{figure} [htbp]\centering
 \includegraphics [width=8cm]{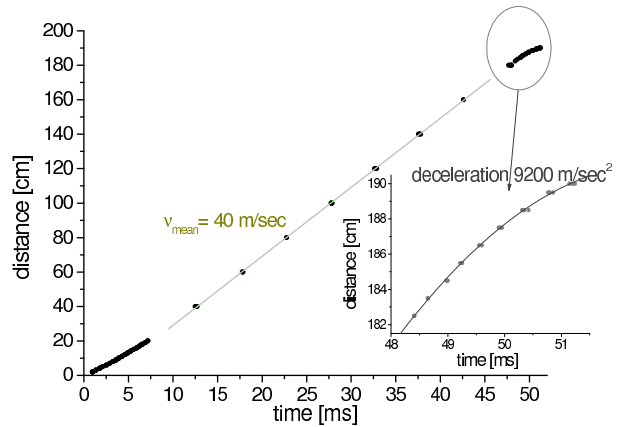}
 \caption{Position versus time  along the whole shuttle tube, and, in the inset, in the deceleration segment. A parabolic fit of the deceleration data leads to an acceleration estimate of about 9200~m/s$^2$. The fit demonstrates the effectiveness of the pneumatic deceleration in minimizing the shock of the final collision.
 \label{fig:speeds}}%
\end{figure}

The time to reach different positions along the tube, was analyzed to evaluate limiting factors and typical acceleration levels. The position versus time is plotted in Fig.\ref{fig:speeds} and shows that an abrupt acceleration (not recorded) occurs in the first centimeters and continues for about 5~ms, during which the sample moves about 10~cm. A constant speed motion (steady state) at a velocity of $v_{\mbox{sample}}\cong$ 40~m/s follows. In this time interval the  short air connection pipeline (M)  constitutes an air-flux bottleneck. 
This is consistent with a Darcy-Weisbach \cite{nevers_book_70} estimate of the pressure loss: 
\begin{equation}
\Delta p= \lambda \rho \frac{l}{D}\frac{v^2}{2},
\label{eq:deltap}
\end {equation}
comparable to the reservoir pressure. Here $\lambda\cong 0.02$ is the friction coefficient, $\rho\cong 7$kg/m$^3$ is the density of the compressed air, $D$ and $l$ are the pipeline diameter and length, respectively, and $v\approx v_{\mbox{sample}}d_{\mbox{int}}^2/D^2$ is the air velocity in the pipeline. Thus shorter displacement times could be achieved by shortening the M connection or --more effectively-- using a larger diameter $(D)$ for that pipeline. An experimental check, performed by increasing the length $l$ of the pipeline  nevertheless showed that  other friction phenomena (most likely involving air layer separating the cartridge surface from the tube wall)  also limit the steady state velocity. In fact doubling $l$ only results in an 8\% reduction of the velocity: much less than the 41\% decrease that would be expected from Eq.\ref{eq:deltap}.

Finally, the pneumatic deceleration occurring in the last part of the tube after the side hole W is clearly evident. The plot shows its effectiveness in reducing the shock of the final collision. The deceleration level inferred from the fit is  $9.2 \times 10^3$~m/sec$^2$. With a constant deceleration by this amount, the cartridge velocity of 40~m/s is reduced to zero within a 9~cm displacement (matching the WB distance) in 4.3~ms.

We tested several kinds of materials for both the cartridges and commercially available tubes, obtaining the best results with a Plexiglass tube and Nylon cartridges. Other choices gave worse results in terms of cartridge abrasion (with polyvinyl chloride), surface roughness and section ellipticity (polyethylene tubes).  Of course further optimization could be achieved by investigating both the materials and production techniques (roughness and ellipticity) of the tube. Besides aliphatic polyamides (Nylon) \cite{weitekamp_prl_83}, other  good candidates for cartridge material are, for instance, the polyoxymethylene (Delrin) \cite{wagner_jmr_99} and polychlorotrifluoroethylene (kel-F) \cite{wu_jmr_95}, all of which were  used successfully in the  similar setups cited, having been chosen for their mechanical strength.

In spite of the reduced shocks, some effort was required to produce rechargeable cartridges suitable for shuttling liquid samples without leaking. To this end, a robust screw-sealed design, like the one illustrated in the inset of Fig.\ref{fig:setup} was developed. 

As mentioned above, it is important to select shuttle tubes with negligible ellipticity. In fact,  several attempts failed due to this kind of imperfection, which rendered the transfer time unreproducible and the whole system insufficiently reliable, possibly also due to  imperfections  in the cartridge shape and to its  unavoidable and additive rotations. Interestingly, in spite of the fact that no constraints were present to prevent cartridge rotations, we verified that the amount of such rotations is negligible in individual displacements, which is an important issue for certain types of measurement.

\section {Halbach assembly}
The  shuttle described is completed by a device for sample magnetization. We designed and built a cost-effective magnetic assembly suitable for applying a strong and homogeneous premagnetization field, while keeping stray fields and their inhomogeneities at an acceptable level in the detection region, whose distance is determined by the shuttle tube length.

\begin{figure}[htbp] \centering
 \includegraphics [width=8cm]{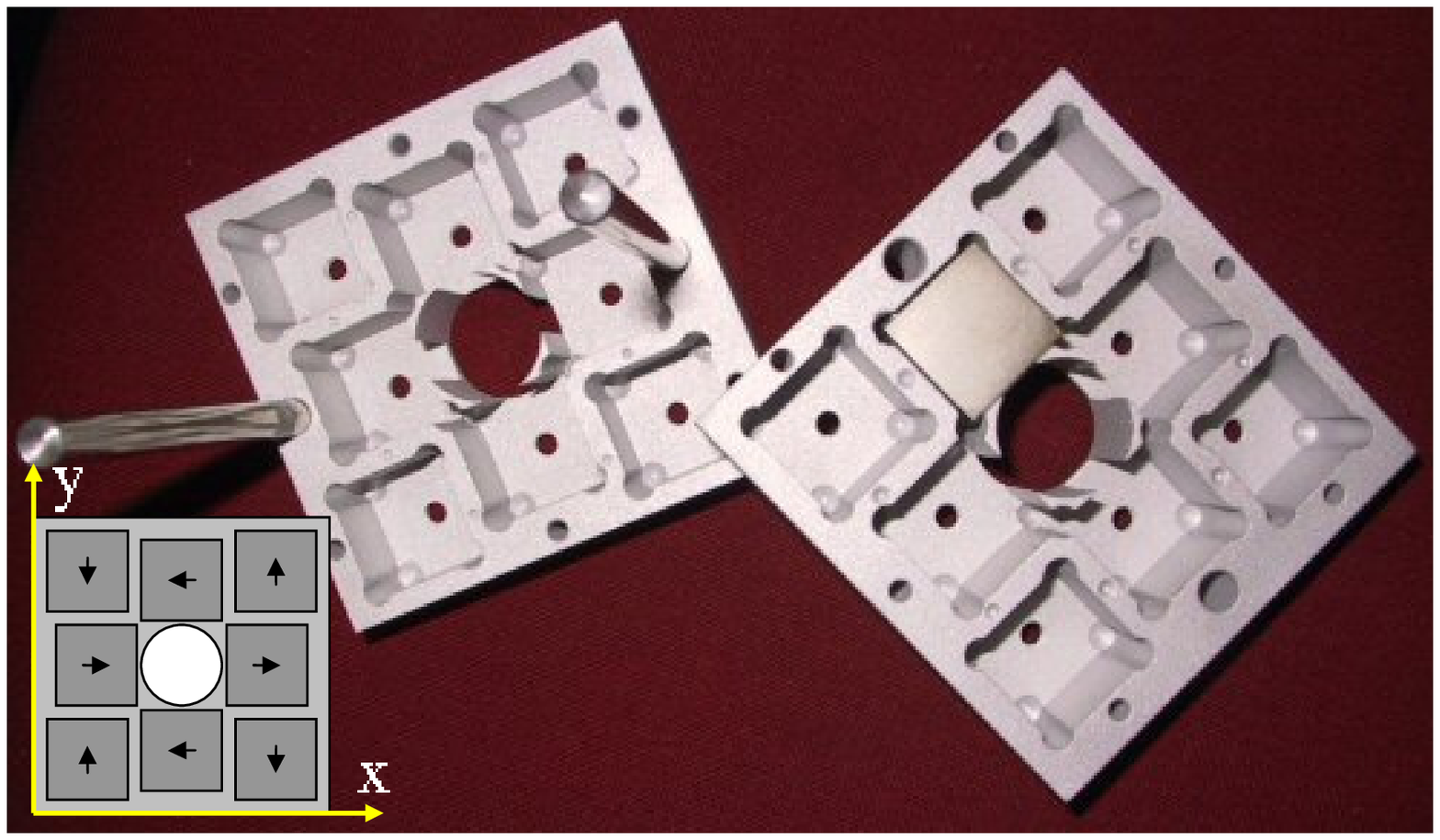}
 \caption{Aluminium shells hosting the magnets, with one magnet inserted. The inset represents the orientation of the  cubic magnets in one layer, which maximizes the field in the hole, while minimizing the field in the external region. The two shells are coupled in such way as to create a cylindrical hole of 25~mm in diameter and 50~mm in length, in which end A of the Plexiglass tube is inserted. 
 \label{fig:shells}}%
\end{figure}

The assembly is based on standard, commercially available Nd magnets. It is sized and shaped in such way as to host  end A of the shuttle tube, having a central hole that matches the external diameter of the shuttle tube. A complete illustration of the shells is available among the supplemental material.
Fig.\ref{fig:shells} shows the aluminium shells used to assemble the magnets in a Halbach configuration. The assembly consists of a two-layer array, each made of 8 cubic magnets, measuring 1~inch per side. 
The field in the middle of the central hole (25~mm in diameter, 50~mm in length) was measured as  1.0~T, in accordance with the value estimated based on the strength of the magnets, whose nominal magnetization level is 1.4~T/$\mu_0$.

The field along the hole axis (let this direction, i.e. the shuttle tube axis, be $\hat z$)  only has the transverse (radial) components, which decay according to a quadrupolar law. The measured value of the component parallel to the cube side (see the inset in Fig.\ref{fig:shells}) has an amplitude $B_x(z)=A/x^4$ with $A=26.3$~T~cm$^4$. The resulting inhomogeneity at a distance of 2~m  is $\partial B_x / \partial z = $~328~pT/cm: a value that is easy to  compensate \cite{belfi_josa_07} and certainly acceptable for  magnetometer operation.


%
%

%


\begin{acknowledgments}
This work was partially supported by the national project FIRB RBAP11ZJFA -005, financed by the Italian Ministry for Education, University and Research. The authors thank  E.~Thorley of Language Box (Siena) for revising the English in the manuscript.
\end{acknowledgments}


\begin{thebibliography}{17}%
\makeatletter
\providecommand \@ifxundefined [1]{%
 \@ifx{#1\undefined}
}%
\providecommand \@ifnum [1]{%
 \ifnum #1\expandafter \@firstoftwo
 \else \expandafter \@secondoftwo
 \fi
}%
\providecommand \@ifx [1]{%
 \ifx #1\expandafter \@firstoftwo
 \else \expandafter \@secondoftwo
 \fi
}%
\providecommand \natexlab [1]{#1}%
\providecommand \enquote  [1]{``#1''}%
\providecommand \bibnamefont  [1]{#1}%
\providecommand \bibfnamefont [1]{#1}%
\providecommand \citenamefont [1]{#1}%
\providecommand \href@noop [0]{\@secondoftwo}%
\providecommand \href [0]{\begingroup \@sanitize@url \@href}%
\providecommand \@href[1]{\@@startlink{#1}\@@href}%
\providecommand \@@href[1]{\endgroup#1\@@endlink}%
\providecommand \@sanitize@url [0]{\catcode `\\12\catcode `\$12\catcode
  `\&12\catcode `\#12\catcode `\^12\catcode `\_12\catcode `\%12\relax}%
\providecommand \@@startlink[1]{}%
\providecommand \@@endlink[0]{}%
\providecommand \url  [0]{\begingroup\@sanitize@url \@url }%
\providecommand \@url [1]{\endgroup\@href {#1}{\urlprefix }}%
\providecommand \urlprefix  [0]{URL }%
\providecommand \Eprint [0]{\href }%
\providecommand \doibase [0]{http://dx.doi.org/}%
\providecommand \selectlanguage [0]{\@gobble}%
\providecommand \bibinfo  [0]{\@secondoftwo}%
\providecommand \bibfield  [0]{\@secondoftwo}%
\providecommand \translation [1]{[#1]}%
\providecommand \BibitemOpen [0]{}%
\providecommand \bibitemStop [0]{}%
\providecommand \bibitemNoStop [0]{.\EOS\space}%
\providecommand \EOS [0]{\spacefactor3000\relax}%
\providecommand \BibitemShut  [1]{\csname bibitem#1\endcsname}%
\let\auto@bib@innerbib\@empty
\bibitem [{\citenamefont {{Kimmich, R.}}\ and\ \citenamefont {{Anoardo,
  E.}}(2004)}]{kimmich_pnmrs_04}%
  \BibitemOpen
  \bibfield  {author} {\bibinfo {author} {\bibnamefont {{Kimmich, R.}}}\ and\
  \bibinfo {author} {\bibnamefont {{Anoardo, E.}}},\ }\href {\doibase
  http://dx.doi.org/10.1016/j.pnmrs.2004.03.002} {\bibfield  {journal}
  {\bibinfo  {journal} {Progress in Nuclear Magnetic Resonance Spectroscopy}\
  }\textbf {\bibinfo {volume} {44}},\ \bibinfo {pages} {257 } (\bibinfo {year}
  {2004})}\BibitemShut {NoStop}%
\bibitem [{\citenamefont {Swanson}\ and\ \citenamefont
  {Kennedy}(1993)}]{swanson_jmr_93}%
  \BibitemOpen
  \bibfield  {author} {\bibinfo {author} {\bibfnamefont {S.}~\bibnamefont
  {Swanson}}\ and\ \bibinfo {author} {\bibfnamefont {S.}~\bibnamefont
  {Kennedy}},\ }\href {\doibase http://dx.doi.org/10.1006/jmra.1993.1121}
  {\bibfield  {journal} {\bibinfo  {journal} {Journal of Magnetic Resonance,
  Series A}\ }\textbf {\bibinfo {volume} {102}},\ \bibinfo {pages} {375 }
  (\bibinfo {year} {1993})}\BibitemShut {NoStop}%
\bibitem [{\citenamefont {Wagner}\ \emph {et~al.}(1999)\citenamefont {Wagner},
  \citenamefont {Dinesen}, \citenamefont {Rayner},\ and\ \citenamefont
  {Bryant}}]{wagner_jmr_99}%
  \BibitemOpen
  \bibfield  {author} {\bibinfo {author} {\bibfnamefont {S.}~\bibnamefont
  {Wagner}}, \bibinfo {author} {\bibfnamefont {T.~R.}\ \bibnamefont {Dinesen}},
  \bibinfo {author} {\bibfnamefont {T.}~\bibnamefont {Rayner}}, \ and\ \bibinfo
  {author} {\bibfnamefont {R.~G.}\ \bibnamefont {Bryant}},\ }\href {\doibase
  http://dx.doi.org/10.1006/jmre.1999.1811} {\bibfield  {journal} {\bibinfo
  {journal} {Journal of Magnetic Resonance}\ }\textbf {\bibinfo {volume}
  {140}},\ \bibinfo {pages} {172 } (\bibinfo {year} {1999})}\BibitemShut
  {NoStop}%
\bibitem [{\citenamefont {Reese}\ \emph {et~al.}(2008)\citenamefont {Reese},
  \citenamefont {Lennartz}, \citenamefont {Marquardsen}, \citenamefont
  {Hoefer}, \citenamefont {Tavernier}, \citenamefont {Carl}, \citenamefont
  {Schippmann}, \citenamefont {Bennati}, \citenamefont {Carlomagno},
  \citenamefont {Engelke},\ and\ \citenamefont {Griesinger}}]{reese_amr_08}%
  \BibitemOpen
  \bibfield  {author} {\bibinfo {author} {\bibfnamefont {M.}~\bibnamefont
  {Reese}}, \bibinfo {author} {\bibfnamefont {D.}~\bibnamefont {Lennartz}},
  \bibinfo {author} {\bibfnamefont {T.}~\bibnamefont {Marquardsen}}, \bibinfo
  {author} {\bibfnamefont {P.}~\bibnamefont {Hoefer}}, \bibinfo {author}
  {\bibfnamefont {A.}~\bibnamefont {Tavernier}}, \bibinfo {author}
  {\bibfnamefont {P.}~\bibnamefont {Carl}}, \bibinfo {author} {\bibfnamefont
  {T.}~\bibnamefont {Schippmann}}, \bibinfo {author} {\bibfnamefont
  {M.}~\bibnamefont {Bennati}}, \bibinfo {author} {\bibfnamefont
  {T.}~\bibnamefont {Carlomagno}}, \bibinfo {author} {\bibfnamefont
  {F.}~\bibnamefont {Engelke}}, \ and\ \bibinfo {author} {\bibfnamefont
  {C.}~\bibnamefont {Griesinger}},\ }\href {\doibase 10.1007/s00723-008-0131-7}
  {\bibfield  {journal} {\bibinfo  {journal} {Applied Magnetic Resonance}\
  }\textbf {\bibinfo {volume} {34}},\ \bibinfo {pages} {301} (\bibinfo {year}
  {2008})}\BibitemShut {NoStop}%
\bibitem [{\citenamefont {Krahn}\ \emph {et~al.}(2010)\citenamefont {Krahn},
  \citenamefont {Lottmann}, \citenamefont {Marquardsen}, \citenamefont
  {Tavernier}, \citenamefont {Turke}, \citenamefont {Reese}, \citenamefont
  {Leonov}, \citenamefont {Bennati}, \citenamefont {Hoefer}, \citenamefont
  {Engelke},\ and\ \citenamefont {Griesinger}}]{krahn_pccp_10}%
  \BibitemOpen
  \bibfield  {author} {\bibinfo {author} {\bibfnamefont {A.}~\bibnamefont
  {Krahn}}, \bibinfo {author} {\bibfnamefont {P.}~\bibnamefont {Lottmann}},
  \bibinfo {author} {\bibfnamefont {T.}~\bibnamefont {Marquardsen}}, \bibinfo
  {author} {\bibfnamefont {A.}~\bibnamefont {Tavernier}}, \bibinfo {author}
  {\bibfnamefont {M.-T.}\ \bibnamefont {Turke}}, \bibinfo {author}
  {\bibfnamefont {M.}~\bibnamefont {Reese}}, \bibinfo {author} {\bibfnamefont
  {A.}~\bibnamefont {Leonov}}, \bibinfo {author} {\bibfnamefont
  {M.}~\bibnamefont {Bennati}}, \bibinfo {author} {\bibfnamefont
  {P.}~\bibnamefont {Hoefer}}, \bibinfo {author} {\bibfnamefont
  {F.}~\bibnamefont {Engelke}}, \ and\ \bibinfo {author} {\bibfnamefont
  {C.}~\bibnamefont {Griesinger}},\ }\href {\doibase 10.1039/C003381B}
  {\bibfield  {journal} {\bibinfo  {journal} {Phys. Chem. Chem. Phys.}\
  }\textbf {\bibinfo {volume} {12}},\ \bibinfo {pages} {5830} (\bibinfo {year}
  {2010})}\BibitemShut {NoStop}%
\bibitem [{\citenamefont {Weitekamp}\ \emph {et~al.}(1983)\citenamefont
  {Weitekamp}, \citenamefont {Bielecki}, \citenamefont {Zax}, \citenamefont
  {Zilm},\ and\ \citenamefont {Pines}}]{weitekamp_prl_83}%
  \BibitemOpen
  \bibfield  {author} {\bibinfo {author} {\bibfnamefont {D.~P.}\ \bibnamefont
  {Weitekamp}}, \bibinfo {author} {\bibfnamefont {A.}~\bibnamefont {Bielecki}},
  \bibinfo {author} {\bibfnamefont {D.}~\bibnamefont {Zax}}, \bibinfo {author}
  {\bibfnamefont {K.}~\bibnamefont {Zilm}}, \ and\ \bibinfo {author}
  {\bibfnamefont {A.}~\bibnamefont {Pines}},\ }\href {\doibase
  10.1103/PhysRevLett.50.1807} {\bibfield  {journal} {\bibinfo  {journal}
  {Phys. Rev. Lett.}\ }\textbf {\bibinfo {volume} {50}},\ \bibinfo {pages}
  {1807} (\bibinfo {year} {1983})}\BibitemShut {NoStop}%
\bibitem [{\citenamefont {Bielecki}\ \emph {et~al.}(1986)\citenamefont
  {Bielecki}, \citenamefont {Zax}, \citenamefont {Zilm},\ and\ \citenamefont
  {Pines}}]{bielecki_rsi_86}%
  \BibitemOpen
  \bibfield  {author} {\bibinfo {author} {\bibfnamefont {A.}~\bibnamefont
  {Bielecki}}, \bibinfo {author} {\bibfnamefont {D.~B.}\ \bibnamefont {Zax}},
  \bibinfo {author} {\bibfnamefont {K.~W.}\ \bibnamefont {Zilm}}, \ and\
  \bibinfo {author} {\bibfnamefont {A.}~\bibnamefont {Pines}},\ }\href@noop {}
  {\bibfield  {journal} {\bibinfo  {journal} {Review of Scientific
  Instruments}\ }\textbf {\bibinfo {volume} {57}} (\bibinfo {year}
  {1986})}\BibitemShut {NoStop}%
\bibitem [{\citenamefont {{Kerwood}}\ and\ \citenamefont
  {{Bolton}}(1987)}]{kerwood_jmr_87}%
  \BibitemOpen
  \bibfield  {author} {\bibinfo {author} {\bibfnamefont {D.}~\bibnamefont
  {{Kerwood}}}\ and\ \bibinfo {author} {\bibfnamefont {F.}~\bibnamefont
  {{Bolton}}},\ }\href@noop {} {\bibfield  {journal} {\bibinfo  {journal}
  {Journal of Magnetic Resonance}\ }\textbf {\bibinfo {volume} {75}},\ \bibinfo
  {pages} {142 } (\bibinfo {year} {1987})}\BibitemShut {NoStop}%
\bibitem [{\citenamefont {Bevilacqua}\ \emph {et~al.}(2013)\citenamefont
  {Bevilacqua}, \citenamefont {Biancalana}, \citenamefont {Dancheva},\ and\
  \citenamefont {Moi}}]{bevilacqua_arnmrs_13}%
  \BibitemOpen
  \bibfield  {author} {\bibinfo {author} {\bibfnamefont {G.}~\bibnamefont
  {Bevilacqua}}, \bibinfo {author} {\bibfnamefont {V.}~\bibnamefont
  {Biancalana}}, \bibinfo {author} {\bibfnamefont {Y.}~\bibnamefont
  {Dancheva}}, \ and\ \bibinfo {author} {\bibfnamefont {L.}~\bibnamefont
  {Moi}},\ }\href {\doibase
  http://dx.doi.org/10.1016/B978-0-12-404716-7.00003-1} {\emph {\bibinfo
  {title} {Annual Report on NMR Spectroscopy}}},\ edited by\ \bibinfo {editor}
  {\bibfnamefont {G.~A.}\ \bibnamefont {Webb}},\ Vol.~\bibinfo {volume} {78}\
  (\bibinfo  {publisher} {Academic Press},\ \bibinfo {year} {2013})\ pp.\
  \bibinfo {pages} {103 -- 148}\BibitemShut {NoStop}%
\bibitem [{\citenamefont {{Ismail}}(2010)}]{ismail_jammc_10}%
  \BibitemOpen
  \bibfield  {author} {\bibinfo {author} {\bibfnamefont {S.~S.}\ \bibnamefont
  {{Ismail}}},\ }\href {\doibase 110.1155/2010/389374} {\bibfield  {journal}
  {\bibinfo  {journal} {Journal of Automated Methods and Management in
  Chemistry}\ }\textbf {\bibinfo {volume} {2010}},\ \bibinfo {pages} {389}
  (\bibinfo {year} {2010})}\BibitemShut {NoStop}%
\bibitem [{\citenamefont {{Bevilacqua}}\ \emph {et~al.}(2009)\citenamefont
  {{Bevilacqua}}, \citenamefont {{Biancalana}}, \citenamefont {{Dancheva}},\
  and\ \citenamefont {{Moi}}}]{bevilacqua_jmr_09}%
  \BibitemOpen
  \bibfield  {author} {\bibinfo {author} {\bibfnamefont {G.}~\bibnamefont
  {{Bevilacqua}}}, \bibinfo {author} {\bibfnamefont {V.}~\bibnamefont
  {{Biancalana}}}, \bibinfo {author} {\bibfnamefont {Y.}~\bibnamefont
  {{Dancheva}}}, \ and\ \bibinfo {author} {\bibfnamefont {L.}~\bibnamefont
  {{Moi}}},\ }\href {\doibase 10.1016/j.jmr.2009.09.013} {\bibfield  {journal}
  {\bibinfo  {journal} {Journal of Magnetic Resonance}\ }\textbf {\bibinfo
  {volume} {201}},\ \bibinfo {pages} {222} (\bibinfo {year}
  {2009})}\BibitemShut {NoStop}%
\bibitem [{\citenamefont {{Bevilacqua}}\ \emph {et~al.}(2012)\citenamefont
  {{Bevilacqua}}, \citenamefont {{Biancalana}}, \citenamefont {{Dancheva}},\
  and\ \citenamefont {{Moi}}}]{bevilacqua_pra_12}%
  \BibitemOpen
  \bibfield  {author} {\bibinfo {author} {\bibfnamefont {G.}~\bibnamefont
  {{Bevilacqua}}}, \bibinfo {author} {\bibfnamefont {V.}~\bibnamefont
  {{Biancalana}}}, \bibinfo {author} {\bibfnamefont {Y.}~\bibnamefont
  {{Dancheva}}}, \ and\ \bibinfo {author} {\bibfnamefont {L.}~\bibnamefont
  {{Moi}}},\ }\href {\doibase 10.1103/PhysRevA.85.042510} {\bibfield  {journal}
  {\bibinfo  {journal} {Physical Review A}\ }\textbf {\bibinfo {volume} {85}},\
  \bibinfo {eid} {042510} (\bibinfo {year} {2012})}\BibitemShut {NoStop}%
\bibitem [{\citenamefont {{Belfi}}\ \emph {et~al.}(2009)\citenamefont
  {{Belfi}}, \citenamefont {{Bevilacqua}}, \citenamefont {{Biancalana}},
  \citenamefont {{Cartaleva}}, \citenamefont {{Dancheva}}, \citenamefont
  {{Khanbekyan}},\ and\ \citenamefont {{Moi}}}]{belfi_josa_09}%
  \BibitemOpen
  \bibfield  {author} {\bibinfo {author} {\bibfnamefont {J.}~\bibnamefont
  {{Belfi}}}, \bibinfo {author} {\bibfnamefont {G.}~\bibnamefont
  {{Bevilacqua}}}, \bibinfo {author} {\bibfnamefont {V.}~\bibnamefont
  {{Biancalana}}}, \bibinfo {author} {\bibfnamefont {S.}~\bibnamefont
  {{Cartaleva}}}, \bibinfo {author} {\bibfnamefont {Y.}~\bibnamefont
  {{Dancheva}}}, \bibinfo {author} {\bibfnamefont {K.}~\bibnamefont
  {{Khanbekyan}}}, \ and\ \bibinfo {author} {\bibfnamefont {L.}~\bibnamefont
  {{Moi}}},\ }\href {\doibase 10.1364/JOSAB.26.000910} {\bibfield  {journal}
  {\bibinfo  {journal} {Journal of the Optical Society of America B Optical
  Physics}\ }\textbf {\bibinfo {volume} {26}},\ \bibinfo {pages} {910}
  (\bibinfo {year} {2009})}\BibitemShut {NoStop}%
\bibitem [{\citenamefont {{Belfi}}\ \emph {et~al.}(2010)\citenamefont
  {{Belfi}}, \citenamefont {{Bevilacqua}}, \citenamefont {{Biancalana}},
  \citenamefont {{Cecchi}}, \citenamefont {{Dancheva}},\ and\ \citenamefont
  {{Moi}}}]{belfi_rsi_10}%
  \BibitemOpen
  \bibfield  {author} {\bibinfo {author} {\bibfnamefont {J.}~\bibnamefont
  {{Belfi}}}, \bibinfo {author} {\bibfnamefont {G.}~\bibnamefont
  {{Bevilacqua}}}, \bibinfo {author} {\bibfnamefont {V.}~\bibnamefont
  {{Biancalana}}}, \bibinfo {author} {\bibfnamefont {R.}~\bibnamefont
  {{Cecchi}}}, \bibinfo {author} {\bibfnamefont {Y.}~\bibnamefont
  {{Dancheva}}}, \ and\ \bibinfo {author} {\bibfnamefont {L.}~\bibnamefont
  {{Moi}}},\ }\href {\doibase 10.1063/1.3441980} {\bibfield  {journal}
  {\bibinfo  {journal} {Review of Scientific Instruments}\ }\textbf {\bibinfo
  {volume} {81}},\ \bibinfo {pages} {065103} (\bibinfo {year}
  {2010})}\BibitemShut {NoStop}%
\bibitem [{\citenamefont {Nevers}(1970)}]{nevers_book_70}%
  \BibitemOpen
  \bibfield  {author} {\bibinfo {author} {\bibfnamefont {N.}~\bibnamefont
  {Nevers}},\ }\href@noop {} {\emph {\bibinfo {title} {Fluid mechanics}}}\
  (\bibinfo  {publisher} {Addison-Wesley Pub. Co},\ \bibinfo {address}
  {Reading, Mass},\ \bibinfo {year} {1970})\BibitemShut {NoStop}%
\bibitem [{\citenamefont {{Donghui Wu}}\ and\ \citenamefont {{Charles S.
  Johnson Jr.}}(1995)}]{wu_jmr_95}%
  \BibitemOpen
  \bibfield  {author} {\bibinfo {author} {\bibnamefont {{Donghui Wu}}}\ and\
  \bibinfo {author} {\bibnamefont {{Charles S. Johnson Jr.}}},\ }\href
  {\doibase http://dx.doi.org/10.1006/jmra.1995.0020} {\bibfield  {journal}
  {\bibinfo  {journal} {Journal of Magnetic Resonance, Series A}\ }\textbf
  {\bibinfo {volume} {116}},\ \bibinfo {pages} {270 } (\bibinfo {year}
  {1995})}\BibitemShut {NoStop}%
\bibitem [{\citenamefont {{Belfi}}\ \emph {et~al.}(2007)\citenamefont
  {{Belfi}}, \citenamefont {{Bevilacqua}}, \citenamefont {{Biancalana}},
  \citenamefont {{Cartaleva}}, \citenamefont {{Dancheva}},\ and\ \citenamefont
  {{Moi}}}]{belfi_josa_07}%
  \BibitemOpen
  \bibfield  {author} {\bibinfo {author} {\bibfnamefont {J.}~\bibnamefont
  {{Belfi}}}, \bibinfo {author} {\bibfnamefont {G.}~\bibnamefont
  {{Bevilacqua}}}, \bibinfo {author} {\bibfnamefont {V.}~\bibnamefont
  {{Biancalana}}}, \bibinfo {author} {\bibfnamefont {S.}~\bibnamefont
  {{Cartaleva}}}, \bibinfo {author} {\bibfnamefont {Y.}~\bibnamefont
  {{Dancheva}}}, \ and\ \bibinfo {author} {\bibfnamefont {L.}~\bibnamefont
  {{Moi}}},\ }\href {\doibase 10.1364/JOSAB.24.002357} {\bibfield  {journal}
  {\bibinfo  {journal} {Journal of the Optical Society of America B Optical
  Physics}\ }\textbf {\bibinfo {volume} {24}},\ \bibinfo {pages} {2357}
  (\bibinfo {year} {2007})}\BibitemShut {NoStop}%
\end{thebibliography}

%

\hrule

\begin{figure} \centering
 \includegraphics [width=15cm]{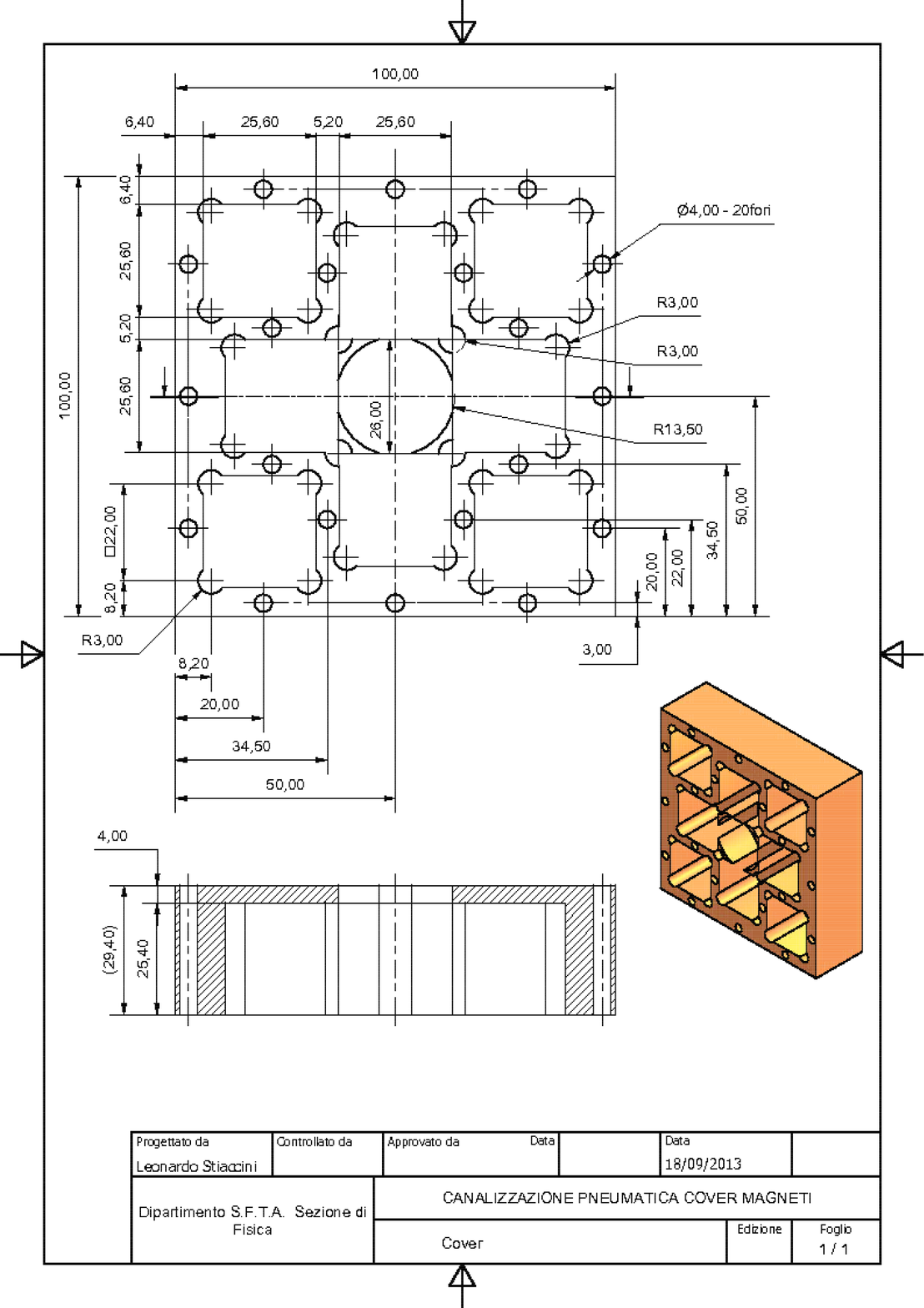}
 \caption{Supplemental material 1/2: drawings of aluminium shells.
        \label{suppl1}}
\end{figure}

\begin{figure} \centering
 \includegraphics [width=15cm]{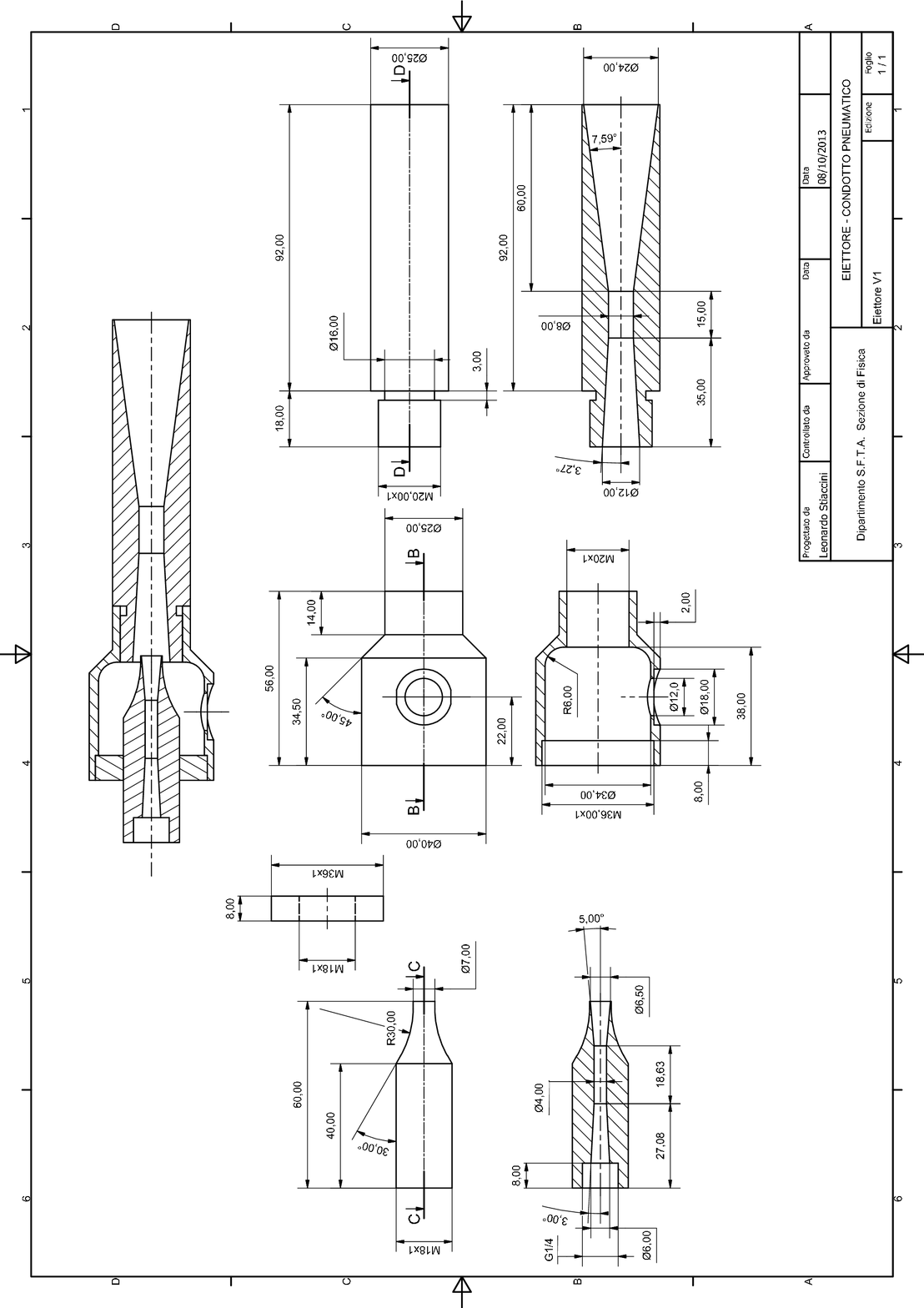}
 \caption{Supplemental material 2/2: drawings of the ejector.
        \label{suppl2}}
\end{figure}

{\footnotesize In the next two pages, Figs.\ref{suppl1} and \ref{suppl2}: supplemental	material.}

\end{document}